\documentclass[%
 reprint,
superscriptaddress,
preprintnumbers,
 amsmath,amssymb,
 aps,
prd,
floatfix,
longbibliography
]{revtex4-2}
\usepackage{amsfonts}
\usepackage{slashed}
\usepackage{graphicx}
\usepackage{dcolumn}
\usepackage{bm}
\usepackage[braket,qm]{qcircuit}
\usepackage{float}
\usepackage[dvipsnames]{xcolor}
\usepackage[pdftex,
            pdftitle={Computing the Mass Shift of Wilson and Staggered Fermions in the Lattice Schwinger Model with Matrix Product States},
            pdfauthor={Takis Angelides, Lena Funcke, Karl Jansen, Stefan Kuehn},
            bookmarks,
            colorlinks,
            linkcolor=myblue,
            citecolor=mymagenta,
            menucolor=black,
            urlcolor=myblue,
            plainpages=false,
            pdfpagelabels,
            hypertexnames=false]{hyperref}
\definecolor{mymagenta}{RGB}{200, 0, 100}
\definecolor{myblue}{RGB}{45, 48, 146}

\begin{document}

\title{Computing the Mass Shift of Wilson and Staggered Fermions in the Lattice Schwinger Model with Matrix Product States}

\author{Takis Angelides}
\affiliation{
 Institut für Physik, Humboldt-Universität zu Berlin, Newtonstr. 15, 12489 Berlin, Germany
}
\affiliation{
 Deutsches Elektronen-Synchrotron DESY, Platanenallee 6, 15738 Zeuthen, Germany
}
\author{Lena Funcke}
\affiliation{
 Transdisciplinary Research Area ``Building Blocks of Matter and Fundamental Interactions'' (TRA Matter) and Helmholtz Institute for Radiation and Nuclear Physics (HISKP), University of Bonn, Nußallee 14-16, 53115 Bonn, Germany
}
\affiliation{
 Center for Theoretical Physics, Co-Design Center for Quantum Advantage, and NSF AI Institute for Artificial Intelligence and Fundamental Interactions, Massachusetts Institute of Technology, 77 Massachusetts Avenue, Cambridge, MA 02139, USA
}
\author{Karl Jansen}
\affiliation{
 Deutsches Elektronen-Synchrotron DESY, Platanenallee 6, 15738 Zeuthen, Germany
}
\affiliation{
 Computation-Based Science and Technology Research Center, The Cyprus Institute, 20 Kavafi Street,
2121 Nicosia, Cyprus
}
\author{Stefan K\"uhn}
\affiliation{
 Deutsches Elektronen-Synchrotron DESY, Platanenallee 6, 15738 Zeuthen, Germany
}
\affiliation{
 Computation-Based Science and Technology Research Center, The Cyprus Institute, 20 Kavafi Street,
2121 Nicosia, Cyprus
}

\date{\today}

\begin{abstract}
Simulations of lattice gauge theories with tensor networks and quantum computing have so far mainly focused on staggered fermions. In this paper, we use matrix product states to study Wilson fermions in the Hamiltonian formulation and present a novel method to determine the additive mass renormalization. Focusing on the single-flavor Schwinger model as a benchmark model, we investigate the regime of a non-vanishing topological $\theta$-term, which is inaccessible to conventional Monte Carlo methods. We systematically explore the dependence of the mass shift on the volume, the lattice spacing, the $\theta$-parameter, and the Wilson parameter. This allows us to follow lines of constant renormalized mass, and therefore to substantially improve the continuum extrapolation of the mass gap and electric field density. For small values of the mass, our continuum results agree with the theoretical prediction from mass perturbation theory. Going beyond Wilson fermions, our technique can also be applied to staggered fermions, and we demonstrate that the results of our approach agree with a recent theoretical prediction for the mass shift at sufficiently large volumes.
\end{abstract}

\maketitle

\section{Introduction}
\label{sec:Introduction}

Lattice gauge theory (LGT) is an essential tool for exploring gauge theories in the non-perturbative regime~\cite{lQCD_gattringer}. After discretizing the Lagrangian on a space-time lattice, stochastic Monte Carlo (MC) methods can be applied to numerically study mass spectra~\cite{Durr2008}, phase diagrams~\cite{Fukushima2011}, and many other static properties. However, standard MC methods suffer from the sign problem in certain parameter regimes, which prevents the investigation of many interesting problems. Prominent examples are the phase diagram of QCD at high baryon chemical potential or QCD in the presence of a topological $\theta$-term~\cite{finite_density_lqcd_1}. In contrast, methods based on the Hamiltonian formulation do not suffer from the sign problem. In particular, tensor network approaches in the Hamiltonian formulation have successfully demonstrated calculations in regimes where conventional MC methods suffer from the sign problem (see, e.g., Refs.~\cite{p1, p2, p3, p4, p5, topo_vac,Funcke2021,Funcke2023,sign_problem_vs_tn,Zache:2021ggw} and Ref.~\cite{practical_intro_mps} for a review). Moreover, in recent years, quantum computing has emerged as a promising new approach for tackling gauge theories in the Hamiltonian formulation (see, e.g., Refs.~\cite{Martinez2016,Kokail2018,Klco2018,Klco2019,Ciavarella2021,Mazzola2021,Funcke:2022uwc,Halimeh:2022pkw} and Ref.~\cite{Funcke2023a} for a review). 

The naive lattice discretization of theories with fermionic degrees of freedom suffers from the fermion doubling problem~\cite{lQCD_gattringer}. The most common approach for avoiding the doubling problem in simulations based on tensor networks and quantum computing has been the usage of Kogut-Susskind staggered fermions~\cite{kogut_susskind_staggered}. While these fermions are easy to implement, they do not allow for fully removing the doublers in 2+1 and 3+1 space-time dimensions. Wilson fermions offer an alternative discretization scheme for avoiding the doublers, by giving them a mass proportional to the inverse lattice spacing~\cite{wilson_fermions_original_work, Creutz1994}. This renders the doublers infinitely heavy when taking the continuum limit, and they completely decouple from the theory. Wilson fermions generalize straightforwardly to any number of space-time dimensions, and allow for fully removing the doublers. However, one complication arising from Wilson fermions is the explicit breaking of axial symmetry, which causes an additive mass renormalization~\cite{Rothe:1992nt}. Methods for computing this mass shift have so far only been proposed and implemented in the Lagrangian formulation.

In this paper, we extend the work done in Ref.~\cite{Angelides2022} and present a method to compute the mass shift in the Hamiltonian formulation, focusing on the Schwinger model with a topological $\theta$-term as a benchmark model. The knowledge of the mass shift allows for following lines of constant renormalized mass. We demonstrate that incorporating the mass shift improves the convergence of the continuum extrapolation and allows for obtaining more precise results with a given amount of resources.

This paper is organized as follows. In Sec.~\ref{sec:theory}, we review the continuum formulation and the lattice formulation of the Schwinger model with a topological $\theta$-term. In Sec.~\ref{sec:methods}, we explain the specific kind of tensor network, the matrix product states (MPS), which we use for our numerical calculations, as well as the new method for determining the mass shift. Subsequently, we present our results for the mass shift and its dependence on the volume, the lattice spacing, the $\theta$-parameter, and the Wilson parameter in Sec.~\ref{sec:results}. We also perform the continuum extrapolations for the electric field density and the Schwinger boson mass after incorporating the mass shift. Finally, we provide a conclusion and an outlook in Sec.~\ref{sec:conclusion}. In Appendix~\ref{appendix:fujikawa}, we provide a short review of the quantum
anomaly equation and the resulting $\theta$-independence of observables in the case of vanishing fermion mass. Since our new method of computing the mass shift is not restricted to a specific type of fermion discretization, we demonstrate our method for staggered fermions in Appendix~\ref{appendix:staggered} and compare our numerical results to the recent theoretical prediction in Ref.~\cite{chiral_dempsey_staggered}. Related work that has made use of the mass shift prediction for staggered fermions in Ref.~\cite{chiral_dempsey_staggered} can be found in Ref.~\cite{staggered_ref_1, staggered_ref_2}.

\section{Theory}
\label{sec:theory}
In this section, we will first review the continuum version of the Schwinger model with a topological $\theta$-term, including expressions for the Schwinger boson mass and the vacuum expectation value of the electric field density. Secondly, we will present the corresponding lattice model with Wilson fermions in the Hamiltonian formulation. 

\subsection{Continuum Schwinger Model}
\label{sec:continuum_theory}

The Schwinger model describes quantum electrodynamics in 1+1 dimensions. In the continuum, the corresponding Hamiltonian density for a single fermion flavor with a topological $\theta$-term is given by
\begin{equation}
\label{hamiltonian_density}
    \mathcal{H} = -i\overline{\psi}\gamma^1\left(\partial_1 - igA_1\right)\psi + m\overline{\psi}\psi + \frac{1}{2}\left(\dot{A}_1 + \frac{g\theta}{2\pi}\right)^2.
\end{equation}
We have chosen temporal gauge, $A_0 = 0$, and therefore only the $F_{01} = \dot{A_1}$ gauge field component appears. The bare coupling of the gauge field to the fermionic field $\psi$ is denoted by $g$, and has units of mass. The fermionic field $\psi$ is a two-component spinor with bare mass $m$. The $\theta$-term $g\theta/2\pi$ in Eq.~\eqref{hamiltonian_density} represents a static background electric field~\cite{tong_gt,adam,coleman_more_about_SM}. The Schwinger model is super-renormalizable; therefore, the bare and renormalized parameters are identical~\cite{super_renormalizability, adam}.

In the massless case, the axial transformation of the fermionic fields, $\psi \to e^{i\gamma_5\alpha}\psi$, is a symmetry of the classical theory but not of the quantum theory. Based on Ref.~\cite{fujikawa_method}, we review in Appendix~\ref{appendix:fujikawa} that this quantum anomaly implies that the $\theta$-parameter becomes unphysical for the massless case and observables should therefore become $\theta$-independent.

For the massive continuum Schwinger model, Ref.~\cite{adam} used mass perturbation theory up to order $\mathcal{O}\left[(m/g)^2\right]$ to derive the following expressions for the vacuum expectation value of the electric field density,
\begin{equation}
\label{adam_efd}
    \frac{\mathcal{F}}{g} = \frac{e^\gamma}{\sqrt{\pi}}\Big(\frac{m}{g}\Big)\sin{\theta} - 8.9139\text{ }\frac{e^{2\gamma}}{4\pi}\Big(\frac{m}{g}\Big)^2\sin{(2\theta)},\\
\end{equation}
and the mass gap, called the Schwinger boson mass,
\begin{align}
\label{adam_mass_gap}
     \frac{M_S}{g} = \frac{1}{\sqrt{\pi}}\Biggr[1 &+ 3.5621\sqrt{\pi}\Big(\frac{m}{g}\Big)\cos{\theta} \\
     &+ \pi\big(5.4807 - 2.0933\cos{(2\theta)}\big)\Big(\frac{m}{g}\Big)^2\Biggr]^{1/2}, \nonumber
\end{align}
where $\gamma = 0.5772156649$ is the Euler-Mascheroni constant. We can see from both expressions that they become independent of $\theta$ when $m/g = 0$, as expected from the axial anomaly. In particular for $m/g = 0$, a non-vanishing background electric field, corresponding to a nonzero $\theta$-term, gets completely screened by fermion-anti-fermion pairs that are created from the vacuum~\cite{coleman_more_about_SM}. This pair creation accumulates negative and positive charges at the ends of the spatial dimension, thereby creating an electric field in the opposite direction to the original background electric field, such that the overall electric field vanishes. The mass gap $M_S/g$, also known as the Schwinger boson mass, represents a stable mesonic bound state of a fermion-anti-fermion pair~\cite{coleman_more_about_SM}.

\subsection{Lattice Formulation with Wilson Fermions}

In order to numerically study the Schwinger model with matrix product states (MPS), we first need a discrete lattice Hamiltonian, which we will introduce in this section. We follow Refs.~\cite{zache, tavernelli_wilson_fermions} and derive the lattice Hamiltonian with Wilson fermions, which yields the Hamiltonian in  Eq.~\eqref{hamiltonian_density} in the continuum limit. Our starting point is the free Dirac Hamiltonian in the continuum,
\begin{equation}
    \label{continuum_free_hamiltonian}
    H^{\text{free}} = \int dx \text{ } \overline{\psi}\left(-i\gamma^1\partial_1 + m\right)\psi.
\end{equation}
To obtain a discretized version of Eq.~\eqref{continuum_free_hamiltonian}, we use the symmetric lattice derivative, $\partial_1 \psi_n = (\psi_{n+1}-\psi_{n-1})/2a$ and replace $\int dx \to a\sum_n$, where $a$ is the lattice spacing and $n$ indicates the lattice site. The resulting Hamiltonian reads
\begin{equation}
\label{discrete_free_hamiltonian}
    H^\text{free}_\text{lat} = a\sum_n \biggr[ m_\text{lat}\overline{\psi}_n\psi_n -i\overline{\psi}_n \gamma^1\left(\frac{\psi_{n+1}-\psi_{n-1}}{2a}\right)\biggr].
\end{equation}
Here, we explicitly distinguish between the continuum mass $m$ and the lattice mass $m_\text{lat}$. The naive lattice Hamiltonian in Eq.~\eqref{discrete_free_hamiltonian} is plagued by fermionic doublers, and we therefore add the Wilson term
\begin{align}
\label{wilson_term}
\begin{split}
    \Delta H_{\rm Wilson} &= -r\frac{a^2}{2}\sum_n\overline{\psi}_n\partial_1^2\psi_n  \\
    &= -r\frac{a^2}{2}\sum_n\overline{\psi}_n\left(\frac{\psi_{n+1}+\psi_{n-1}-2\psi_n}{a^2}\right),
\end{split}
\end{align}
to the Hamiltonian in order to remove them~\cite{wilson_fermions_original_work}. Here, $r$ is the Wilson parameter, and we use the symmetrized discrete second derivative. The Wilson term in Eq.~\eqref{wilson_term} gives a mass proportional to $r/a$ to the doublers, such that they decouple in the continuum limit, $a\to 0$~\cite{wilson_fermions_original_work}.

When adding the Wilson term in Eq.~\eqref{wilson_term} to the Hamiltonian in Eq.~\eqref{discrete_free_hamiltonian}, we obtain the free lattice Dirac Hamiltonian with Wilson fermions,
\begin{align}
\label{free_lattice_wilson_hamiltonian}
    H^\text{free}_\text{lat, Wilson} = \sum_n \biggr[(&am_\text{lat} + r)\overline{\psi}_n\psi_n \nonumber\\
    &-\overline{\psi}_n\left(\frac{r+i\gamma^1}{2}\right)\psi_{n+1}\\
    &+\overline{\psi}_n\left(\frac{-r+i\gamma^1}{2}\right)\psi_{n-1}\biggr]\nonumber.
\end{align}
Next, we would like to gauge the theory, such that $U(1)$ gauge transformations of the form $\psi_n \to e^{i\beta_n}\psi_n$ correspond to a (local) symmetry of the Hamiltonian. To this end, we introduce the link operator $U_n$ which is placed on the link between sites $n$ and $n+1$, and transforms under gauge transformations as $U_n \to e^{i\beta_n}U_ne^{-i\beta_{n+1}}$. The conjugate field to $U_n$ is the electric field $E_n$, which is also acts on the links, and satisfies the commutation relations $[E_{n} , U_{n'}] = g \delta_{n,n'} U_{n'}$. By introducing this gauge symmetry, we obtain the interacting lattice Hamiltonian
\begin{align}
\label{interacting_discrete_hamiltonian}
     H^\text{int}_\text{lat, Wilson} = \sum_n\biggr[&-\overline{\psi}_n\left(\frac{r+i\gamma^1}{2}\right)U_n\psi_{n+1} \nonumber\\
     &+ \overline{\psi}_n\left(\frac{-r+i\gamma^1}{2}\right)U^\dagger_n\psi_{n-1} \\
     &+ (am_\text{lat}+r)\overline{\psi}_n\psi_n + a\frac{E_n^2}{2} \biggr].\nonumber    
\end{align}
For numerical simulations, it is more convenient to work with a dimensionless formulation. Hence, we use dimensionless operators, $L_n = E_n/g$ and $\phi_{n, \alpha} = (-1)^n\sqrt{a}\psi_{n, \alpha}$, where $\alpha$ labels the spinor component, and we consider the dimensionless Hamiltonian $\widetilde{W} = (2/ ag^2)H^\text{int}_\text{lat, Wilson}$.

The physical states of the Hamiltonian have to obey Gauss's law, $\forall n$ $L_n - L_{n-1} = Q_n$, where $Q_n \equiv \phi^\dagger_n\phi_n - 1$ is the charge operator. For open boundary conditions (OBC), the set of constraints can be solved explicitly,
\begin{equation}
    L_n = l_0 + \sum_{k=1}^{n}Q_k,
    \label{l_n_equation}
\end{equation}
where $l_0$ is the electric field value on the left boundary and nothing but a background electric field corresponding to $\theta / 2\pi$ in Eq.~\eqref{hamiltonian_density}. Thus, the fermionic charge content completely determines the flux content of the links after fixing $l_0$. Substituting Eq.~\eqref{l_n_equation} into Eq.~\eqref{interacting_discrete_hamiltonian}, we can eliminate the electric field. Applying the unitary transformation $\phi_n \to \prod_{k < n}U^\dagger_k \phi_n$ to the resulting expression, the gauge field can be fully removed~\cite{hamer97}. The Hamiltonian obtained in this way is directly constrained to the physical subspace and its eigenstates fulfill Gauss's law. 

For convenience in the numerical simulations, we map the fermionic fields to spin operators by choosing the ordering $\phi_{n,\alpha} \to \chi_{2n-2+\alpha}$ and applying a Jordan-Wigner transformation, $\chi_n = \prod_{k<n}(i\sigma^z_k)\sigma_n^-$~\cite{jordan_wigner}. In the previous expression, the matrices $\sigma_n^a$ with $a\in\{x,y,z\}$ are the usual Pauli matrices acting on site $n$, and we define $\sigma^\pm_n \equiv (\sigma_n^x \pm i \sigma^y_n)/2$. The final dimensionless lattice Hamiltonian in the spin formulation, using $\gamma_0 = \sigma^x$ and $\gamma_1 = i\sigma^z$, is given by
\begin{align}
\label{model_hamiltonian_w}
    \widetilde{W} = &\text{ }ix(r-1)\sum_{n=1}^{N-1}\left(\sigma_{2n}^-\sigma_{2n+1}^+ - \sigma_{2n}^+\sigma_{2n+1}^-\right) \nonumber\\
    &+ix(r+1)\sum_{n=1}^{N-1}(\sigma_{2n-1}^+\sigma^z_{2n}\sigma^z_{2n+1}\sigma^-_{2n+2}) \nonumber\\
    &-ix(r+1)\sum_{n=1}^{N-1}(\sigma_{2n-1}^-\sigma^z_{2n}\sigma^z_{2n+1}\sigma^+_{2n+2}
    ) \\
    &+ 2i\left(\frac{m_\text{lat}}{g}\sqrt{x} + xr\right)\sum_{n=1}^N\left(\sigma_{2n-1}^-\sigma_{2n}^+ - \sigma_{2n-1}^+\sigma_{2n}^-\right)  \nonumber\\
    &+\sum_{n=1}^{N-1}\left(l_0 + \sum_{k=1}^n Q_k\right)^2. \nonumber
\end{align}
In the expression above, $x \equiv 1/(ag)^2$ is the inverse lattice spacing squared in units of the coupling. Equation~\eqref{model_hamiltonian_w} describes the Schwinger model with Wilson fermions in the spin formulation on a lattice with dimensionless physical volume $agN = N/\sqrt{x}$.

In the continuum, states which do not have a vanishing total charge have infinite energy~\cite{lattice_schwinger_model}. Hence, only states with zero total charge have finite energy and can be labelled physical. To ensure that our results are in the sector of vanishing total charge, we add a penalty term to the Hamiltonian,   
\begin{align}
    W = \widetilde{W} + \lambda \left(\sum_{n=1}^NQ_n\right)^2,
    \label{eq:final_hamiltonian}
\end{align}
where the constant $\lambda$ has to be chosen large enough. 

The two terms corresponding to the electric field energy and the charge penalty, respectively, can be expressed in terms of Pauli matrices using the expression $Q_k = \left(\sigma^z_{2k-1} + \sigma^z_{2k}\right)/2$ for the charge operator, which yields
\begin{align}
\begin{split}
    &\sum_{n=1}^{N-1}\left(l_0 + \sum_{k=1}^n Q_k\right)^2 + \lambda \left(\sum_{n=1}^NQ_n\right)^2   \\
    & = l_0\sum_{n=1}^{2N-2}\left(N- \left \lceil \frac{n}{2} \right \rceil \right)\sigma^z_n \\
    &\quad + \frac{1}{2}\sum_{n=1}^{2N}\sum_{k=n+1}^{2N}\left(N- \left \lceil \frac{k}{2} \right \rceil  + \lambda\right)\sigma^z_n\sigma^z_k  \\
    &\quad + l_0^2(N-1) +\frac{1}{4}N(N-1) + \frac{\lambda N}{2}, 
\end{split}
\label{eq:energy_charge_penalty}
\end{align}
where $\left\lceil . \right\rceil$ is the ceiling function. Equation~\eqref{eq:energy_charge_penalty} includes long-range interactions originating from the Coulomb force mediated by massless photons. The Hamiltonian in Eq.~\eqref{eq:final_hamiltonian}, after substituting in Eq.~\eqref{eq:energy_charge_penalty}, can be efficiently addressed with MPS, as shown in the next section. 

\section{Methods}
\label{sec:methods}
In this section, we first briefly review the MPS techniques used for computing the ground state and excited states. Then, we propose a new method to compute the mass shift by measuring the mass dependence of the vacuum expectation value of the electric field density.

\subsection{Matrix Product States}
\label{sec:MPS}
Tensor network states are a family of entanglement-based ansätz for the wave function of quantum many-body systems, where the amount of entanglement in the ansatz state is limited by the size of the tensors~\cite{variational_tn_algorithm, schollwock, practical_intro_mps}. A widely used class of one-dimensional tensor network states are the MPS, which for a system of $2N$ spins on a lattice with OBC reads
\begin{equation}
\label{mps_ket_rep}
    \ket{\psi} = \sum_{i_1,i_2,..,i_{2N}} A^{i_1}_{1,\alpha_1}A^{i_2}_{\alpha_1,\alpha_2}...A^{i_{2N}}_{\alpha_{2N-1},1} \ket{i_1} \otimes \ket{i_2} ... \otimes \ket{i_{2N}}.
\end{equation}
In the expression above, the indices $\alpha_k$, commonly referred to as virtual indices, are implicitly contracted. For a fixed value of $i_k$, the $A^{i_k}_{\alpha_{k-1},\alpha_{k}}$ in the ansatz can be interpreted as matrices, hence the name MPS. The maximum value of the virtual indices, corresponding to the maximum size of the matrices $A^{i_k}$, is referred to as the bond dimension $D$ of the MPS. The value of $D$ determines the number of variational parameters in the ansatz and the amount of entanglement that can be present in the MPS. The physical indices $i_n$ range over the physical degrees of freedom on each lattice site $n$. For our case of spin $1/2$, they take two possible values. 

In order to compute the ground state, we adopt a standard variational algorithm successively updating the tensors in order to minimize the energy expectation value $E=\langle\psi|W|\psi\rangle$~\cite{variational_tn_algorithm, schollwock, practical_intro_mps}. For our simulations, we use the implementation in the ITensors Julia package~\cite{iTJulia}. We continue the optimization until the relative change of the energy is below a certain tolerance $\eta$, which we will set to $10^{-12}$ in our simulations. After obtaining the ground state $\ket{\psi_0}$, the first excited state can be computed in a similar fashion by considering the Hamiltonian $W_{\text{eff}} = W + |E_0|\ket{\psi_0}\bra{\psi_0}$~\cite{banuls_mass_spectrum}. Assuming that the first excited state has an energy $E_1<0$~\footnote{This can always be achieved by subtracting a large enough constant from the Hamiltonian}, the ground state of $W_{\text{eff}}$ corresponds to the first excited state of $W$. Having obtained an MPS approximation for a state $\ket{\psi}$, we can proceed to measure observables $O$ by expressing them as matrix product operators~\cite{Pirvu_2010} and contracting the network corresponding to  $\langle\psi|O|\psi\rangle$. 

\subsection{New Method for Computing the Mass Shift}
\label{sec:mass_shift_measurement}

The main goal of our work is to compute the mass shift, which arises due to the Wilson term in Eq.~\eqref{wilson_term}. The technique we propose is not only applicable to Wilson fermions, but also to staggered fermions~\cite{banuls_mass_spectrum}, as we will demonstrate in Appendix~\ref{appendix:staggered}. While the mass shift for staggered fermions in the one-flavor Schwinger model can be analytically computed for periodic boundary conditions (PBC)~\cite{chiral_dempsey_staggered}, the mass shift of Wilson fermions can only be numerically investigated. For this numerical investigation, let us express the renormalized mass $m_r/g$ in terms of the lattice mass $m_{\text{lat}}/g$,
\begin{equation}
\label{mass_shift_general_equation}
       \frac{m_r}{g} = \frac{m_{\text{lat}}}{g} + \text{MS}(N, x, l_0, r).
\end{equation}
Here, MS refers to the mass shift as a function of the parameters $N$, $x$, $l_0$, and $r$ in units of the coupling $g$. 

\begin{figure}[htp!]
    \centering
    \includegraphics[width=\linewidth]{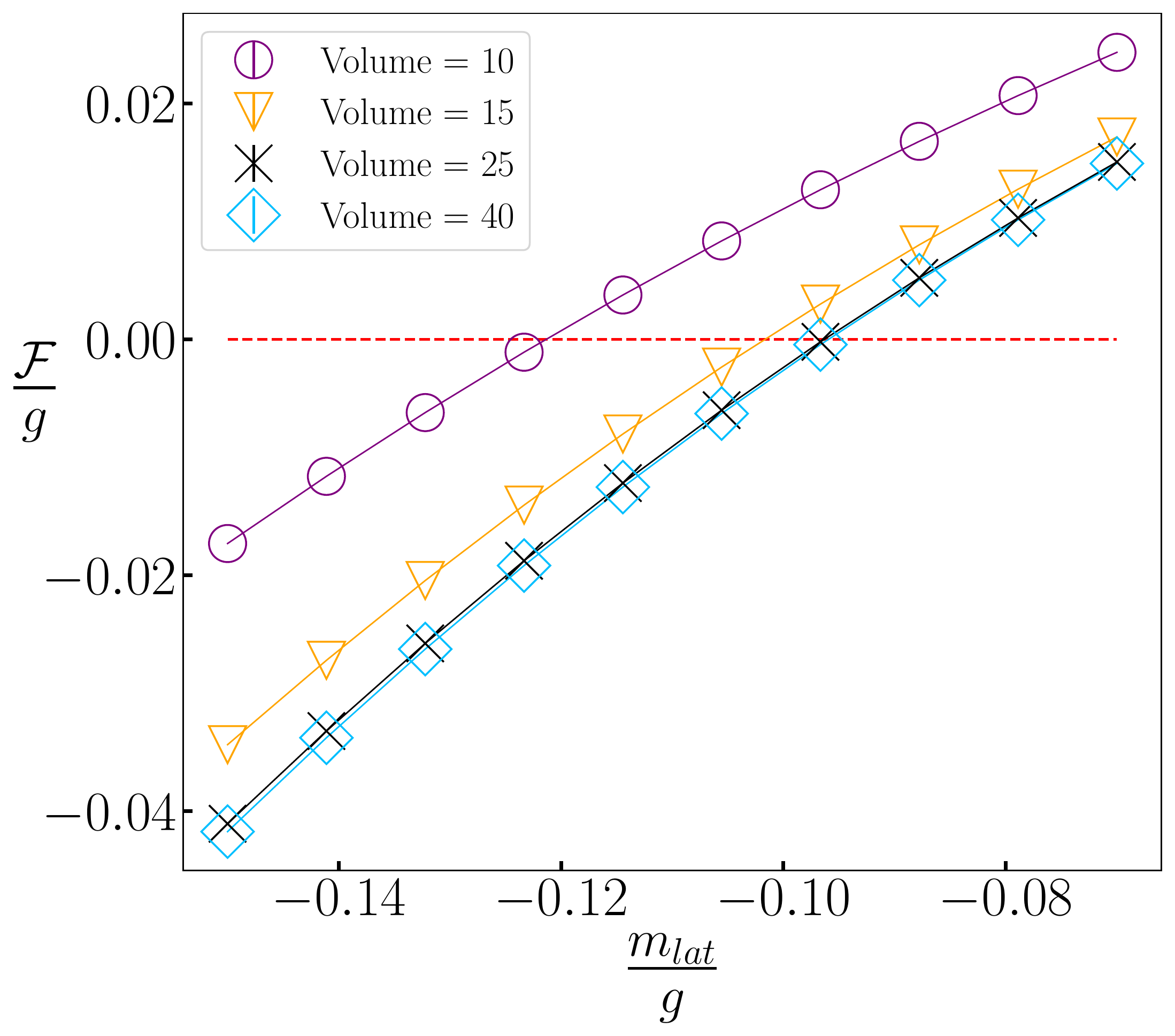}
    \caption{Electric field density $\mathcal{F}/g$ versus lattice mass $m_\text{lat}/g$. The markers represent data for different physical volumes $N/\sqrt{x}=10$ (purple circles), 15 (yellow triangles), 25 (black crosses), and 40 (blue squares), demonstrating the presence of finite-volume effects for the mass shift. Following Eq.~\eqref{adam_efd}, the electric field density vanishes for $m_r/g = 0$; therefore, the intercepts of the data curves with $\mathcal{F}/g = 0$ (red dashed line) correspond to minus the mass shift for a given volume. Note that the error bars are much smaller than the markers and thus are not visible.}
    \label{fig:efd_vs_lattice_mass_different_volumes}
\end{figure}

To obtain the mass shift for either Wilson or staggered fermions, we measure the vacuum expectation value of the dimensionless electric field density $\mathcal{F}/g$ as a function of the lattice mass $m_\text{lat}/g$, see Fig.~\ref{fig:efd_vs_lattice_mass_different_volumes}. Using the fact that $\mathcal{F}/g=0$ for $m_r=0$, we can determine the mass shift from this plot. To this end, we fit a quadratic function to our numerical data, following the analytical continuum prediction in Eq.~\eqref{adam_efd}, and afterwards identify the term $\text{MS}(N, x, l_0, r)$ in Eq.~\eqref{mass_shift_general_equation} with minus the value of $m_\text{lat}/g$ for which $\mathcal{F}/g=0$. This new method to determine the mass shift requires numerical data at negative values of $m_\text{lat}/g$ and therefore cannot be implemented with standard lattice Monte Carlo methods due to the sign problem. The MPS method explained in Sec.~\ref{sec:MPS} circumvents the sign problem, which is the reason why we use this method for measuring $\mathcal{F}/g$ and computing the mass shift.

For measuring $\mathcal{F}/g$ and computing MS, we first calculate $L_n = E_n/g$ from the fermionic charge content using  Eq.~\eqref{l_n_equation}. In order to avoid boundary effects due to OBC, we only keep links from the center of the lattice and average the electric field over those links, which we then identify with $\mathcal{F}/g$. To estimate the errors, we follow Ref.~\cite{topo_vac} and  consider two sources of errors, which we add in quadrature: first, the errors from the extrapolation of $\mathcal{F}/g$ to infinite bond dimension $D$, and second, the errors from the extrapolation of $\mathcal{F}/g$ to the continuum limit, $ag\to 0$. Unlike in Ref.~\cite{topo_vac}, we do not perform an infinite-volume extrapolation and explicitly work at fixed finite volume $N/\sqrt{x}$.

This method for computing the MS cannot be used directly for $\theta = 0$ as $\mathcal{F}/g$ will be zero for all values of mass. However, one can measure the MS for small values of $\theta$ and then extrapolate to $\theta = 0$.

\section{Results}
\label{sec:results}
In this section, we numerically test our new method to determine the mass shift and investigate the dependence of the resulting mass shift on the volume, the lattice spacing, the $\theta$-parameter and the Wilson parameter. Furthermore, we compare the continuum electric field density at fixed finite volume and finite mass to the analytical results from mass perturbation theory. Finally, we show the same comparison for the Schwinger boson mass for zero bare fermion mass. For all our simulations, we use the standard choice for the Wilson parameter, $r=1$, unless stated otherwise. Moreover, we set the strength of the penalty term enforcing vanishing total charge to $\lambda=100$, where we have checked that this strength is sufficient to avoid states with nonzero charge.

\subsection{Dependence of Mass Shift on Parameters}
In the following, we will provide a detailed numerical study of how the mass shift due to the Wilson term depends on the volume $N/\sqrt{x}$, the lattice spacing $ag$, the $\theta$-parameter, and the Wilson parameter $r$.

\subsubsection{Dependence on Volume}
\label{sec:volume}
\begin{figure}[t!]
    \centering
    \includegraphics[width=\linewidth]{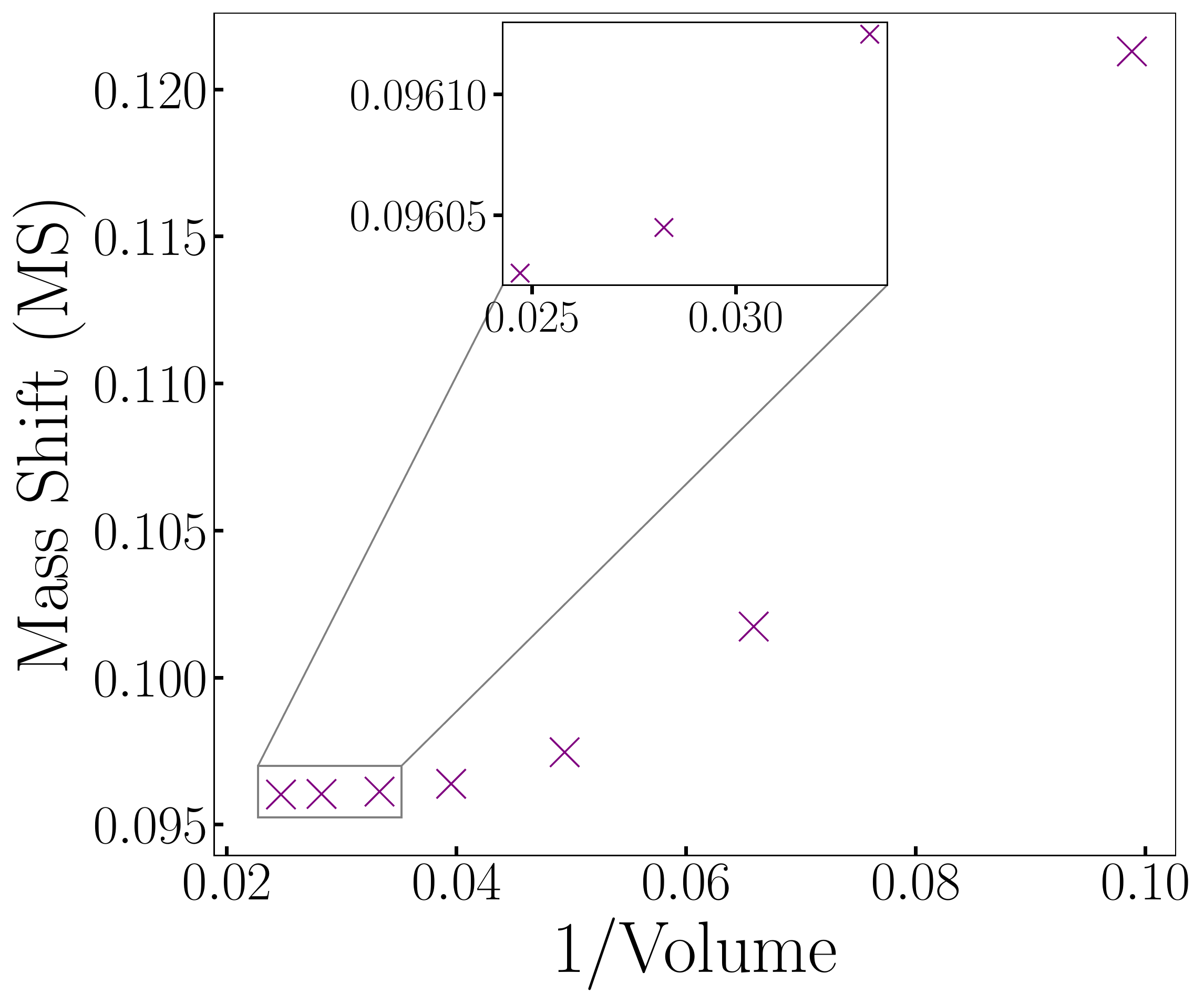}
    \caption{Mass shift (MS) versus inverse volume. The markers show data for volumes $N/\sqrt{x}$ between 10 and 40, where $x = 10$ and $l_0 = 0.1$ are fixed. The MS exhibits a plateau for volumes $N/\sqrt{x}\gtrsim 30$, with a relative difference in the MS of $\sim 0.01$\% for the largest two volumes of 35 and 40 (see inset). As before, the error bars are much smaller than the markers and thus are not visible.}
    \label{fig:mass_shift_vs_inverse_volume}
\end{figure}

To examine the volume dependence of the mass shift, we fix $x = 10$ and $l_0 = 0.1$ and compute MS for different volumes $N/\sqrt{x}$, following the procedure outlined in Sec.~\ref{sec:mass_shift_measurement}. Our results are shown in Fig.~\ref{fig:mass_shift_vs_inverse_volume}. We observe that the mass shift initially shows a strong dependence on the lattice volume, before it eventually plateaus upon reaching a volume of $N/\sqrt{x}\approx 30$. The strong volume dependence of the mass shift for small volumes is likely due to finite-volume effects of the electric field density, as we will explain in the following. 

On a lattice with OBC in the sector of vanishing total charge, the electric field is given by $l_0$ on both ends. Since the electric field in the whole system is determined by the fermionic charge content via Eq.~\eqref{l_n_equation}, and the charge can only take values $\{-1,0,1\}$, a number of links from the right and left boundary are required to realize a bulk value for the electric field differing from $l_0$.  For small volumes, such a bulk region might not be able to form, which results in finite-size effects for the electric field density. In turn, these effects propagate to the mass shift.  

\subsubsection{Dependence on Lattice Spacing}
\label{sec:lattice_spacing_dependence_of_mass_shift}
In order to extrapolate observables to the continuum limit, one needs to evaluate their values at different lattice spacing $ag$ while keeping the renormalized mass $m_r/g$ constant. This requires the knowledge of the mass shift as a function of $ag$. 

\begin{figure}[htp!]
    \centering
    \includegraphics[width=\linewidth]{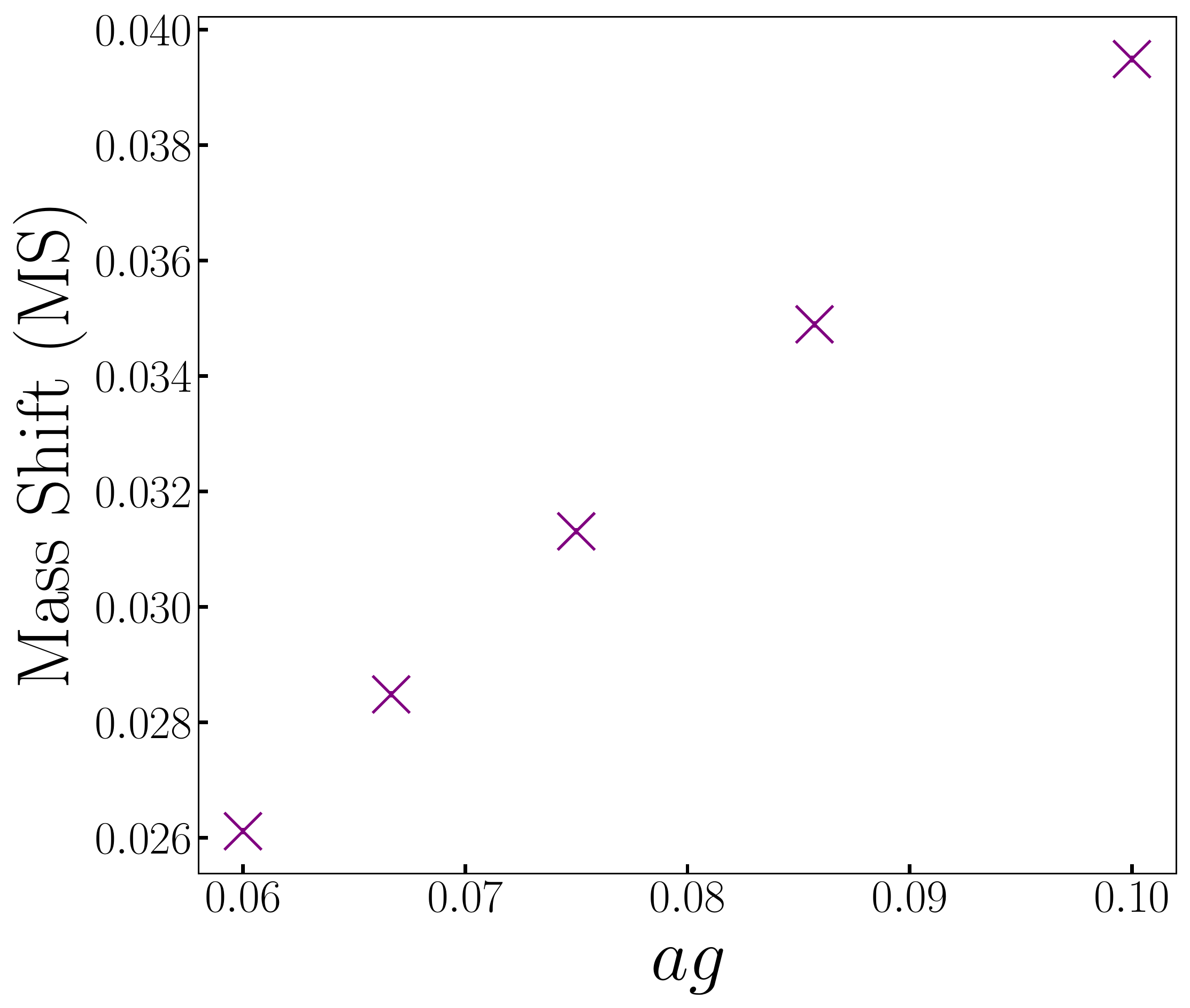}
    \caption{Mass shift (MS) as a function of the lattice spacing $ag$. We fix $l_0 = 0.125$ and $N/\sqrt{x} = 30$, with $N$ ranging from 300 to 500. To first order, the $ag$-dependence of the MS is linear. As before, the error bars are much smaller than the markers and thus are not visible.}
    \label{fig:mass_shift_vs_ag}
\end{figure}

To examine the dependence of the mass shift on the lattice spacing, we fix $l_0 = 0.125$ and set the volume to $N/\sqrt{x} = 30$, which is a value that we have seen to be large enough to avoid noticeable finite-size effects (see Sec.~\ref{sec:volume}). Figure~\ref{fig:mass_shift_vs_ag} shows our results for the mass shift as a function of the lattice spacing. The data reveal that the mass shift depends linearly on $ag$ to first order. In particular, we observe that the mass shift decreases as we approach the continuum limit, as expected.

\subsubsection{Dependence on \texorpdfstring{$\theta$}\text{-parameter}}
\begin{figure}[t!]
    \centering
    \includegraphics[width=\linewidth]{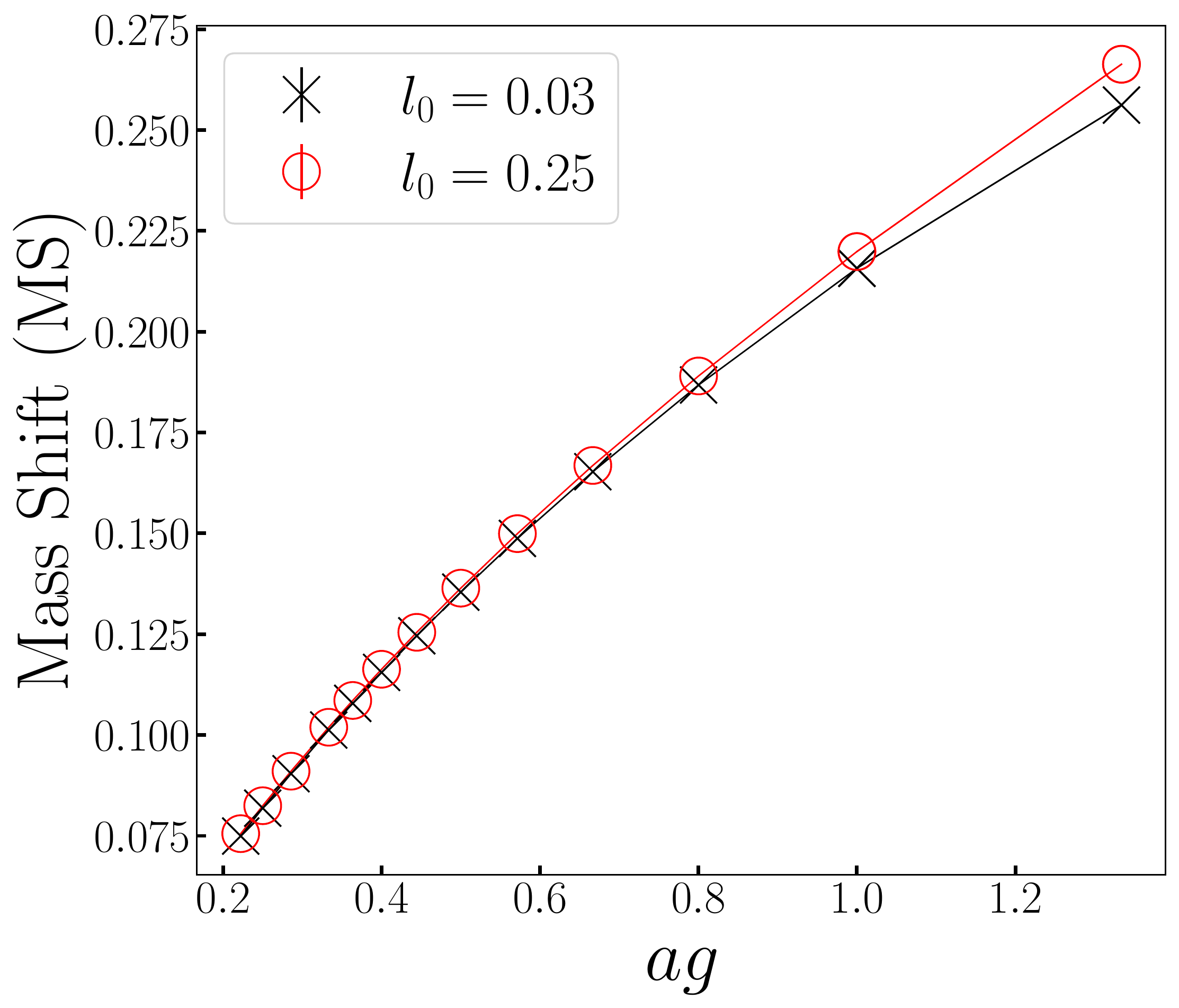}
    \caption{Mass shift (MS) versus lattice spacing $ag= 1/\sqrt{x}$ for two different values of the background field, $l_0 = \theta/(2\pi) = 0.03$ (black crosses) and $0.25$ (red circles), demonstrating that the MS is different when $l_0 = \theta/2\pi$ varies. The volume is fixed to $N/\sqrt{x} = 20$,  with $N$ ranging from 25 to 90. As before, the error bars are much smaller than the markers and thus are not visible.}
    \label{fig:mass_shift_vs_ag_different_l_0}
\end{figure}

As outlined in Sec.~\ref{sec:continuum_theory}, the $\theta$-parameter becomes unphysical for vanishing fermion mass in the continuum limit. This is due to the axial anomaly, as reviewed in Appendix~\ref{appendix:fujikawa}. However, the axial anomaly is not exact on the lattice, and some remnant dependence on the $\theta$-parameter appears when measuring the mass shift at $m_r/g = 0$~\cite{topo_vac}. This can be directly seen in our numerical data in Fig.~\ref{fig:mass_shift_vs_ag_different_l_0}, which shows the dependence of the mass shift on the lattice spacing $ag$ for two different values of the background field, $l_0=\theta/2\pi$, at a fixed physical volume of $N/\sqrt{x}=20$. 

\begin{figure}[htp!]
    \centering
    \includegraphics[width=\linewidth]{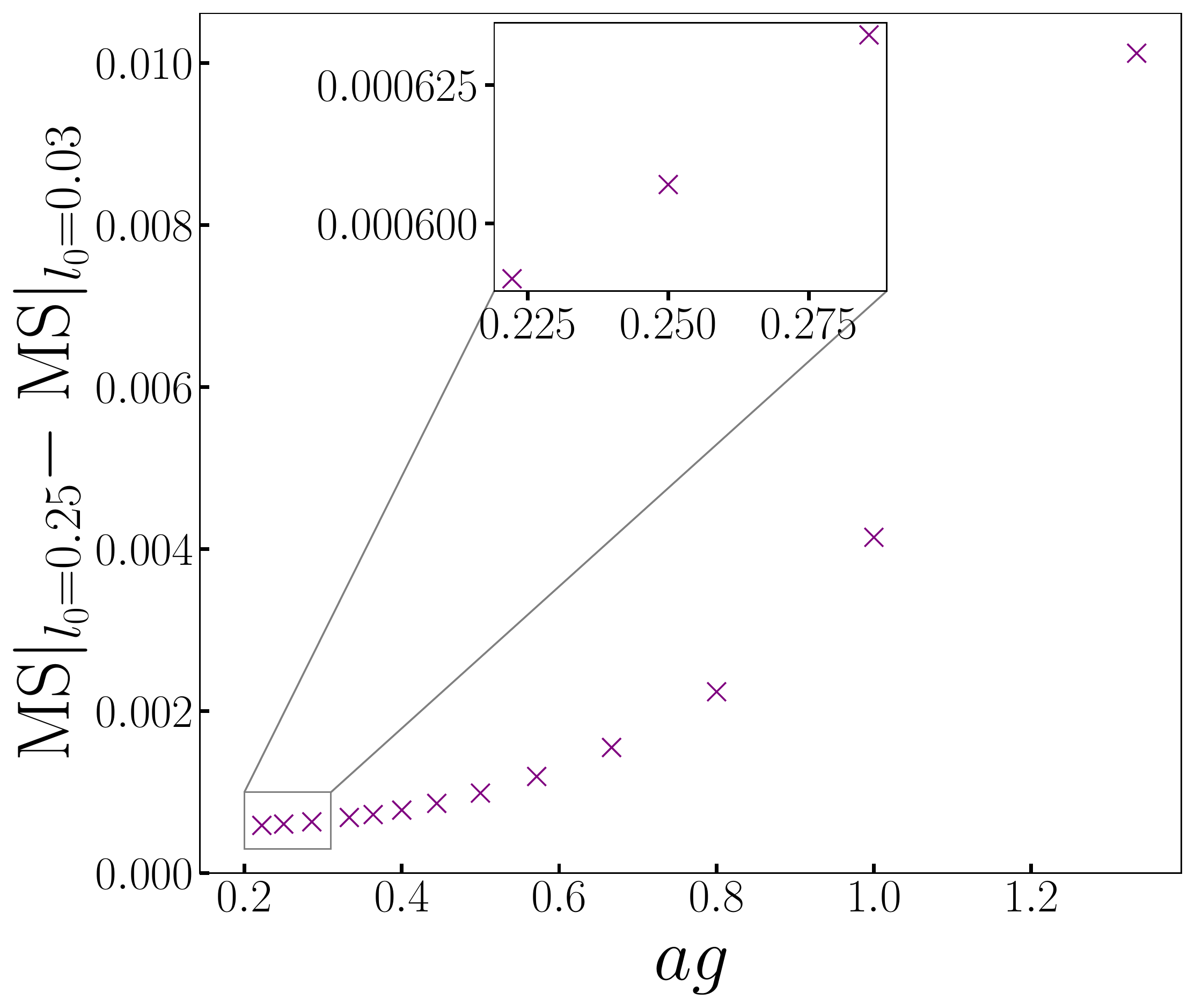}
    \caption{Difference in mass shift (MS) between two different values of the background field, $l_0 = \theta/(2\pi) = 0.25$ and $0.03$, see Eq.~\eqref{eq:DMS}, as a function of the lattice spacing $ag$. The volume is fixed to $N/\sqrt{x} = 20$, with $N$ ranging from 25 to 90. The inset shows data for $x= 1/(ag)^2 = 12.25$, 16, and 20.25, which demonstrate that the $\theta$-dependence of the MS becomes negligible for small $ag$. As before, the error bars are much smaller than the markers and thus are not visible.}
    \label{fig:mass_shift_difference_vs_ag}
\end{figure}

To illustrate the dependence of the mass shift on the $\theta$-parameter in more detail, Fig.~\ref{fig:mass_shift_difference_vs_ag} shows the difference in the mass shift between the two different $l_0$ values, 
\begin{equation}
\Delta\text{MS}\equiv \text{MS}|_{l_0=0.25}-\text{MS}|_{l_0=0.03}.
\label{eq:DMS}
\end{equation}
While we observe a noticeable difference $\Delta\text{MS}$ for large lattice spacing $ag \sim 1$, this difference decreases when decreasing $ag$ and eventually becomes negligible around $ag \lesssim 0.3$. This result agrees with the expectation that the axial anomaly is restored towards the continuum limit. Note that the $l_0$ dependence of the mass shift only vanishes in the infinite-volume limit. Thus, for our finite volume of $N/\sqrt{x}=20$, a small $l_0$ dependence is expected to remain, which agrees with the data in Fig.~\ref{fig:mass_shift_difference_vs_ag}.

\begin{figure}[htp!]
\centering
\includegraphics[width=\linewidth]{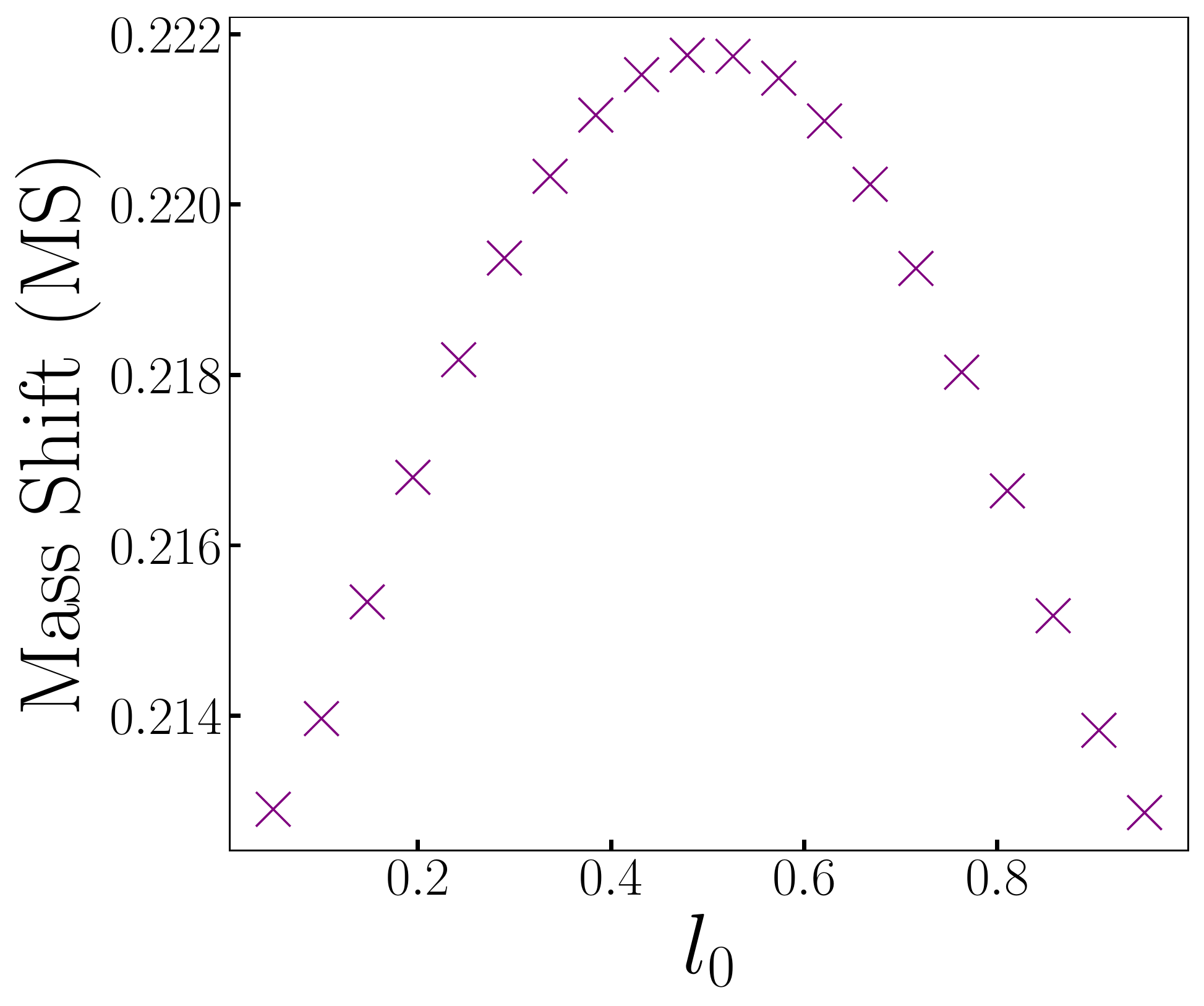}
\caption{Mass shift (MS) versus background electric field $l_0$. The field $l_0 = \theta/2\pi$ is swept over a full period between 0 and 1, and the MS shows the expected periodicity in $l_0$. The data points correspond to $l_0 \in [0.01, 0.9526]$, $N = 100$, and $x = 1$. As before, the error bars are much smaller than the markers and thus are not visible.}
\label{fig:mass_shift_vs_l_0}
\end{figure}

The physics of the Schwinger model is periodic in $\theta$ with period $2\pi$, or, equivalently, periodic in $l_0$ with period 1~\cite{coleman_more_about_SM}, as reviewed in Appendix~\ref{appendix:fujikawa}. To investigate if this periodicity is also reflected in the mass shift, we study the mass shift over a full period of $l_0$ between 0 and 1. As depicted in Fig.~\ref{fig:mass_shift_vs_l_0}, our numerical data for the mass shift shows the expected periodicity. The mass shift increases for $l_0 < 0.5$ upon reaching a peak at 0.5 and then decreases for $l_0 > 0.5$, while being symmetric around $l_0 = 0.5$ or, equivalently, around $\theta = \pi$. 

\subsubsection{Dependence on Wilson Parameter}
\begin{figure}[htp!]
    \centering
    \includegraphics[width=\linewidth]{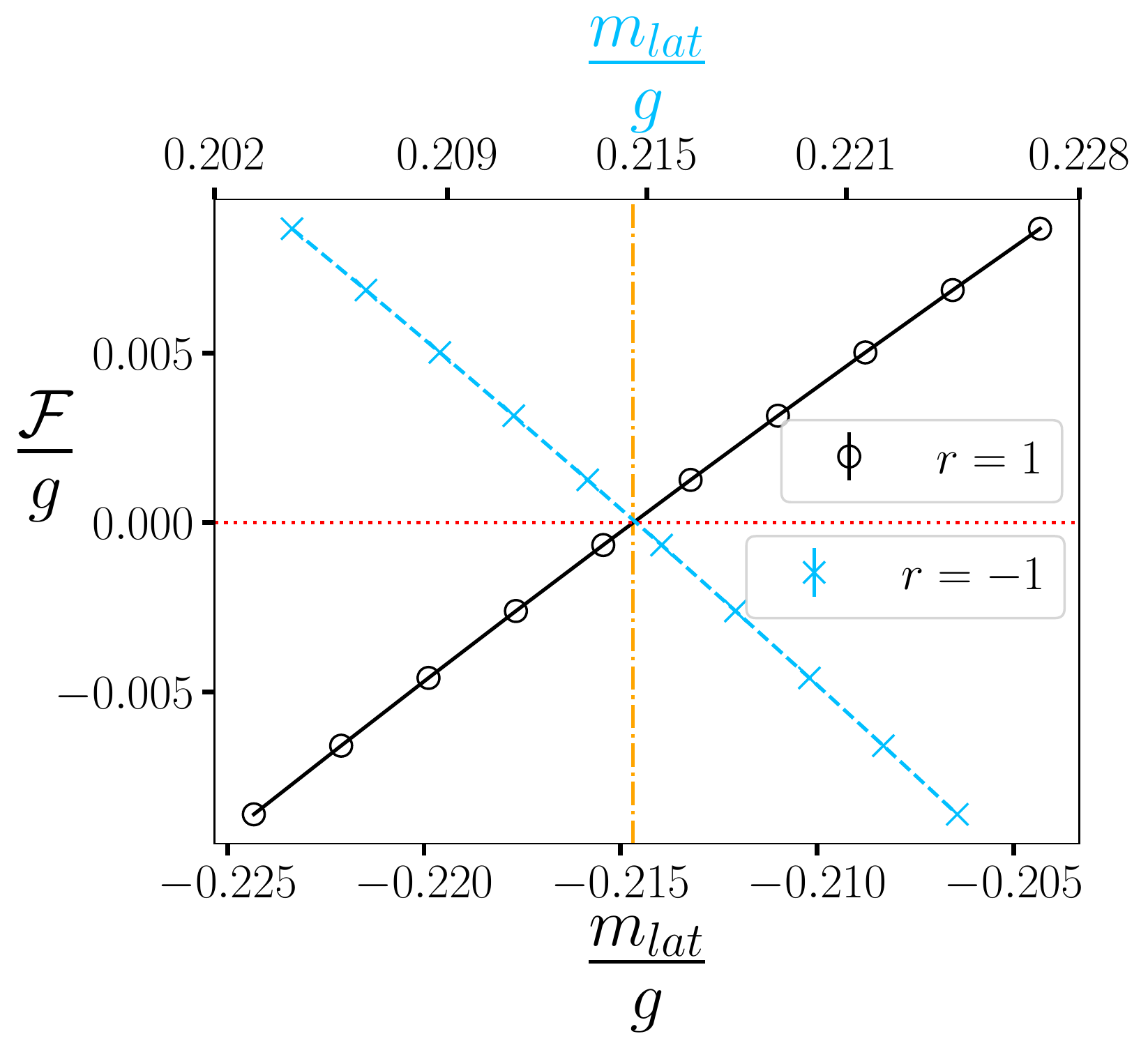}
    \caption{Electric field density $\mathcal{F}/g$ as a function of the lattice mass $m_{\text{lat}}/g$ for the Wilson parameter $r = 1$ (black solid line with lower $x$-axis) and $r = -1$ (blue dashed line with upper $x$-axis). We fix $N = 100$, $x = 1$, and $l_0 = 0.125$. The horizontal red dotted line indicates $\mathcal{F}/g=0$, and the orange dash-dotted vertical line passes the intersection point of the blue and black lines with the red dotted line. The intersection is at a value of $m_\text{lat}/g = 0.214681$ for $r=-1$ (upper $x$-axis) and  $m_\text{lat}/g =-0.214681$ for $r=1$ (lower $x$-axis). As before, the error bars are much smaller than the markers and thus are not visible. }
    \label{fig:efd_vs_lattice_mass_different_r}
\end{figure}

The Hamiltonian in Eq.~\eqref{interacting_discrete_hamiltonian} has a spurious symmetry $\psi \to \gamma_5 \psi$, $m_\text{lat}/g \to -m_\text{lat}/g$, and $r \to -r$, where ``spurious'' means that this symmetry is only present when also transforming the Wilson parameter $r$. Evaluating Eq.~\eqref{mass_shift_general_equation} for $m_r/g = 0$, the symmetry implies that the mass shift is anti-symmetric under $r \to -r$, which yields MS$(N, x, l_0, r) = -$MS$(N, x, l_0, -r)$. With our MPS approach, we can directly check this behavior and study the model for $r=\{-1,1\}$ as well as negative and positive values of $m_\text{lat}/g$, while keeping  $N$, $x$, and $l_0$ fixed. Our numerical results are depicted in Fig.~\ref{fig:efd_vs_lattice_mass_different_r}. We see that the data indeed follow the expected behavior, and that the mass shift obtained for $r = 1$  is equal to the one obtained for $r = -1$ up to a sign.

\subsection{Continuum Extrapolations}
The knowledge of the mass shift allows us to follow lines of constant renormalized mass as we approach the continuum. Thus, it helps to substantially improve the extrapolation of our data to the continuum limit. In the following, we will demonstrate this improvement for two observables: the electric field density and the Schwinger boson mass.

\subsubsection{Electric Field Density}
To examine the mass dependence of the electric field density in the continuum and to compare our numerical data to the perturbative result in Eq.~\eqref{adam_efd}, we choose a fixed volume $N/\sqrt{x} = 20$ and set $l_0 = 0.125$. After computing the mass shift for various values of $ag$, we can extrapolate our numerical lattice data to the continuum while keeping the value of the renormalized mass $m_r/g$ fixed.

\begin{figure}[htp!]
    \centering
    \includegraphics[width=\linewidth]{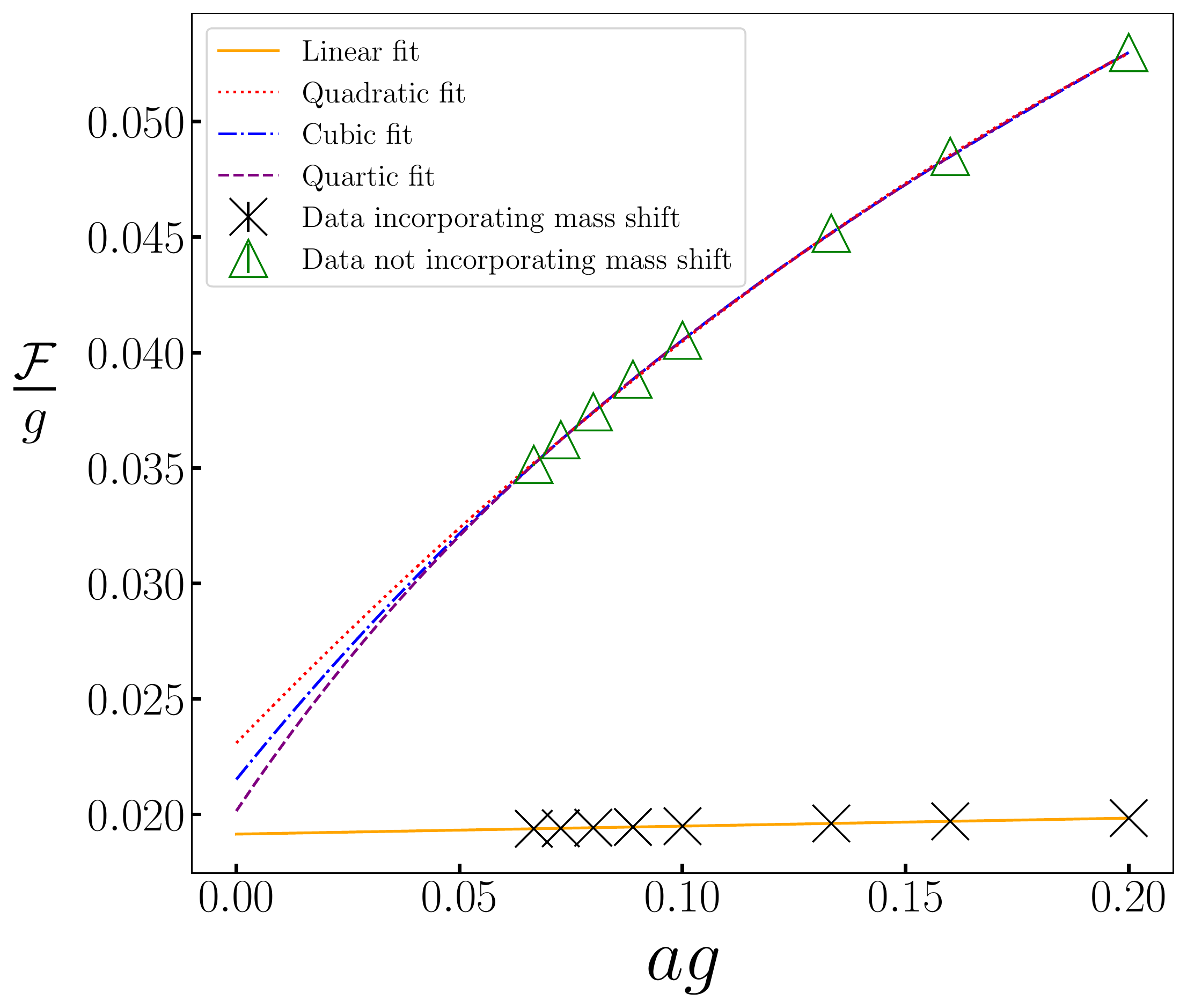}
    \caption{Electric field density $\mathcal{F}/g$ versus lattice spacing $ag$. The markers represent data incorporating the mass shift for $m_r/g=0.03$  (black crosses) and data  without incorporating the mass shift for $m_\text{lat}/g=0.03$ (green triangles). We keep $l_0 = 0.125$ and $N/\sqrt{x} = 20$ fixed. For the data including the mass shift, a linear fit is sufficient, which is expected from the $\mathcal{O}(ag)$ scaling behaviour of non-improved Wilson fermions~\cite{order_a_wislon}. For the data without the mass shift, we approximate $\mathcal{F}/g$ at $ag\to 0$ by fitting a quadratic, cubic, and quartic polynomial and taking a weighted average of the resulting $y$-intercepts. The weights correspond to the mean square error of each fit, and the error on the $y$-intercept for each fit is found following Sec.~\ref{sec:mass_shift_measurement}. We determine the final error by adding the individual errors in weighted quadrature. As before, the error bars are much smaller than the markers and thus are not visible.} 
    \label{fig:efd_vs_ag}
\end{figure}

Figure~\ref{fig:efd_vs_ag} shows an example for the extrapolation procedure, and provides a comparison between the data obtained when incorporating the mass shift and following a line of constant $m_r/g$ (black crosses), as opposed to just setting $m_\text{lat}/g$ to a constant value (green triangles). These data demonstrate the importance of knowing the mass shift for the extrapolation of observables to the continuum. For the data points not incorporating the mass shift (green triangles in Fig.~\ref{fig:efd_vs_ag}), we observe considerable lattice effects, which render the extrapolation to the limit $ag\to 0$ challenging and thus result in large uncertainties for the central value. Focusing on the data points incorporating the mass shift (black crosses in Fig.~\ref{fig:efd_vs_ag}), we see that those are closer to the continuum limit, i.e., the difference between the extrapolated value for $\mathcal{F}/g$ and the data point for the smallest value of $ag$ is significantly smaller.

\begin{figure}[htp!]
    \centering
    \includegraphics[width=\linewidth]{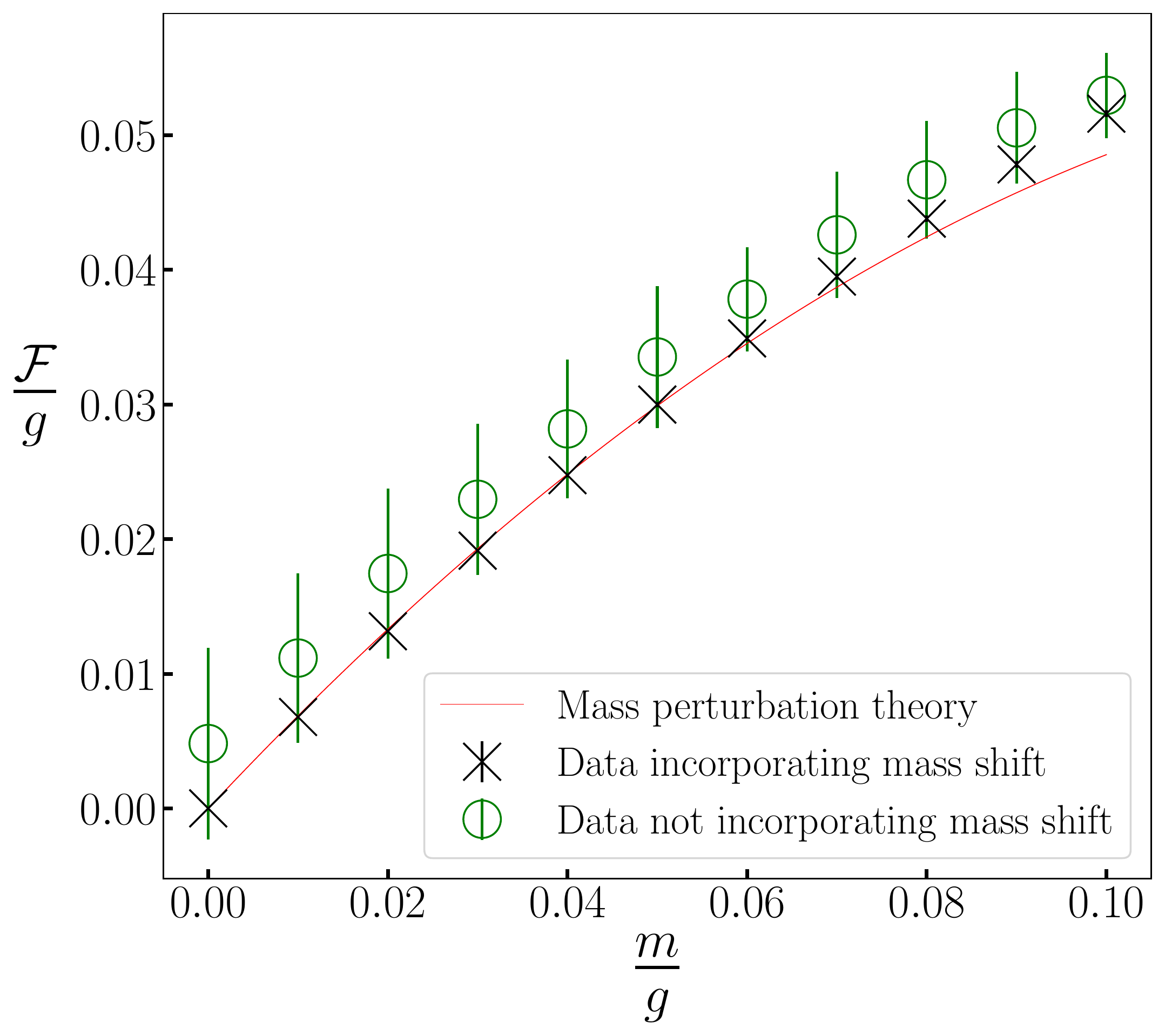}
    \caption{Electric field density $\mathcal{F}/g$ extrapolated to the continuum as a function of the mass $m/g$. For the data incorporating the mass shift (black crosses), $m/g$ corresponds to the renormalized mass $ m_r/g=m_\text{lat}/g$ + MS. For the data without incorporating the mass shift (green circles), $m/g$ is given by $m_\text{lat}/g$. For the continuum prediction from mass perturbation theory (red line), see Eq.~\eqref{adam_efd}, $m/g$ is the continuum mass. For the extrapolation to $ag = 0$, we use eight points with $N \in [100, 300]$ and fixed $l_0 = 0.125$ and  $N/\sqrt{x} = 20$. As before, the error bars for the black crosses are much smaller than the markers and thus are not visible.}
    \label{fig:efd_extrapolated_vs_m_over_g}
\end{figure}

When repeating this procedure for various values of the mass, we can map out the mass dependence of the electric field density in the continuum, which is shown in Fig.~\ref{fig:efd_extrapolated_vs_m_over_g}. Again, we provide data obtained for incorporating the mass shift and data without considering the mass shift. Focusing on the data points without the mass shift, we see that those have considerable error bars throughout the entire range of masses we study, with a tendency to larger errors for smaller values of $m/g$. For small values of $m/g$, the data points agree with the results from mass perturbation theory in Eq.~\eqref{adam_efd} within error bars; however, the central values are consistently above the theoretical prediction.

In contrast, the data points incorporating the mass shift have noticeably smaller error bars, despite using the same lattices and therefore the same numerical resources as for the data without the mass shift. For small values of $m/g$, the data perfectly agree with the perturbative result. As expected, when increasing the values of $m/g$, perturbation theory is eventually no longer suitable to describe the electric field density, and our numerical data differ from the perturbative prediction.

\subsubsection{Schwinger Boson Mass}

\begin{figure}[t!]
    \centering
    \includegraphics[width=\linewidth]{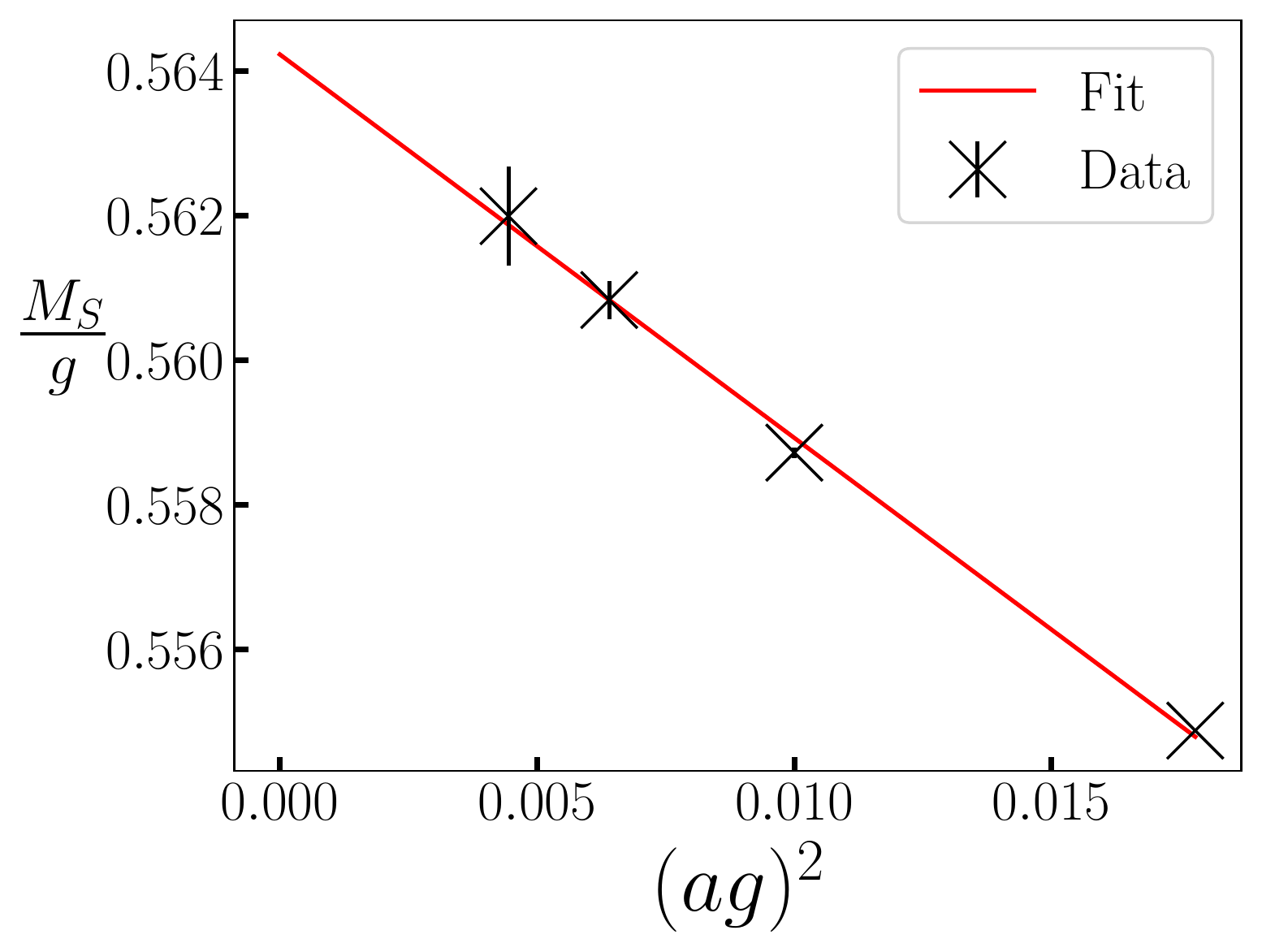}
    \caption{Schwinger boson mass or vector mass gap, $M_S/g$, as a function of the lattice spacing squared, $(ag)^2$. For the data (black crosses), we set $l_0 = 0.125$ and $m_r/g = 0$ and fix the volume to $N/\sqrt{x} = 40$ with $N \in [300, 600]$. For the continuum extrapolation using a quadratic function in $ag$ (red line), we find $M_S/g = 0.5642 \pm 0.0011$, which agrees with the theoretical prediction of $M_S/g = 1/\sqrt{\pi} \approx 0.5641$ in Eq.~\eqref{adam_mass_gap}.}
    \label{fig:mass_gap_vs_ag}
\end{figure}

Finally, we consider the vector mass gap of the theory, which is called the Schwinger boson mass, for the case of $m_r/g = 0$. Using our dimensionless Hamiltonian in Eq.~\eqref{eq:final_hamiltonian} and denoting $\widetilde{W_0}$ and $\widetilde{W_1}$ as the energies of the ground state and the first excited state of $\widetilde{W}$, the Schwinger boson mass in units of the coupling corresponds to~\cite{banuls_mass_spectrum}
\begin{equation}    
    \frac{M_S}{g} = \frac{1}{2\sqrt{x}}(\widetilde{W_1}-\widetilde{W_0}) - \frac{2m_r}{g}.
    \label{schwinger_boson_mass_equation}
\end{equation}
Again, we study this quantity for various values of $ag$ and then we extrapolate to the limit $ag\to 0$, similar to the electric field density. Figure~\ref{fig:mass_gap_vs_ag} shows the continuum extrapolation for the Schwinger boson mass for vanishing renormalized fermion mass and $l_0=0.125$.  

For the extrapolation, we choose a quadratic fit in $ag$ (see the red line in Fig.~\ref{fig:mass_gap_vs_ag}). The reason for this choice is twofold. First, for non-improved Wilson fermions~\cite{order_a_wislon}, we expect that observables such as the electric field density and the energy have $\mathcal{O}(ag)$ effects (see also Fig.~\ref{fig:efd_vs_ag}). Second, due to the dimensionless formulation, Eq.~\eqref{schwinger_boson_mass_equation} introduces another factor of $ag$. Thus, this results in leading-order corrections of $(ag)^2$ for the Schwinger boson mass.  When performing the continuum extrapolation using a quadratic function, we obtain a value of $M_S/g = 0.5642 \pm 0.0011$ for the Schwinger boson mass. Our numerical result is in good agreement with the theoretical calculation in Eq.~\eqref{adam_mass_gap}, which predicts $M_S/g = 1/\sqrt{\pi} \approx 0.5641$.

\section{Conclusion and Outlook}
\label{sec:conclusion}
In this paper, we developed a new method that allows for explicitly determining the additive mass renormalization of Wilson fermions in the Hamiltonian formulation. Focusing on the lattice Schwinger model with a topological $\theta$-term as a benchmark model, our method relies on the fact that the electric field density goes to zero when the renormalized fermion mass vanishes. Of course, when applying the method to other models beyond the Schwinger model, a different observable beyond the electric field density might need to be considered.

For the lattice Schwinger model with a $\theta$-term, we computed the mass shift and studied its dependence on the physical volume, the lattice spacing, the topological $\theta$-parameter, and the Wilson parameter. Our numerical results show that the volume dependence of the mass shift is strong for small volumes but becomes negligible for volumes larger than $N/\sqrt{x} \approx 30$. Moreover, the mass shift is anti-symmetric in the Wilson parameter and strongly depends on the lattice spacing, as expected. The mass shift also shows a weak dependence on the $\theta$-parameter for finite lattice spacing, due to the lattice distortion of the axial anomaly, which becomes negligible as the lattice spacing goes to zero.

Using our results for the mass shift, we were able to follow lines of constant renormalized mass as we approached the continuum. As two examples, we studied the continuum limit of the electric field density and the Schwinger boson mass of the theory. Our results demonstrate that incorporating the mass shift significantly improves the convergence toward the continuum limit as well as the accuracy of the results. For small masses, our numerical data show excellent agreement with results from mass perturbation theory.

Although our study focused on Wilson fermions, our method is not limited to a particular fermion discretization. Therefore, in Appendix~\ref{appendix:staggered}, we demonstrate that the method can also be used to determine the mass shift of staggered fermions. Our numerical data show good agreement with recent theoretical predictions~\cite{chiral_dempsey_staggered}, provided the volume is sufficiently large. 

In both cases of Wilson and staggered fermions, the additive mass renormalization is positive, which implies that our method requires numerical data at negative values of the bare lattice mass $m_\text{lat}/g$. Therefore, the method cannot be implemented with the conventional Monte Carlo approach due to the sign problem, and we employed MPS to compute the electric field density and to determine the mass shift.

Even though we used a tensor network approach, we would like to emphasize that our method is completely general and can be applied to arbitrary Hamiltonian methods, including quantum computing. The electric field density can be readily measured on a quantum device and would allow for similar studies in upcoming quantum computing experiments. Our MPS data can serve as a benchmark for such experiments. 

In this context, we would also like to note that our method will be relevant for the recent proposals to combine small-scale quantum computations with large-scale Monte Carlo simulations of lattice gauge theories, e.g., in order to address the problems of critical slowing down~\cite{Clemente:2022cka} and interpolator optimization~\cite{Avkhadiev:2019niu, Avkhadiev:2022ttx}. These proposals require the implementation of the same fermion discretization, and our method to determine the mass shift of Wilson fermions in the Hamiltonian formulation provides a crucial step into this direction.

\begin{acknowledgments}
This work is partly funded by the European Union’s Horizon 2020 Research and Innovation Programme under the Marie Sklodowska-Curie COFUND scheme with grant agreement no.\ 101034267. 
L.F.\ is partially supported by the U.S.\ Department of Energy, Office of Science, National Quantum Information Science Research Centers, Co-design Center for Quantum Advantage (C$^2$QA) under contract number DE-SC0012704, by the DOE QuantiSED Consortium under subcontract number 675352, by the National Science Foundation under Cooperative Agreement PHY-2019786 (The NSF AI Institute for Artificial Intelligence and Fundamental Interactions, \url{http://iaifi.org/}), and by the U.S.\ Department of Energy, Office of Science, Office of Nuclear Physics under grant contract numbers DE-SC0011090 and DE-SC0021006.
S.K.\ acknowledges financial support from the Cyprus Research and Innovation Foundation under projects ``Future-proofing Scientific Applications for the Supercomputers of Tomorrow (FAST)'', contract no.\ COMPLEMENTARY/0916/0048, and ``Quantum Computing for Lattice Gauge Theories (QC4LGT)'', contract no.\ EXCELLENCE/0421/0019.
This work is funded by the European Union’s Horizon Europe Framework Programme (HORIZON) under the ERA Chair scheme with grant agreement no.\ 101087126.
This work is supported with funds from the Ministry of Science, Research and Culture of the State of Brandenburg within the Centre for Quantum Technologies and Applications (CQTA). 
\end{acknowledgments}

\appendix

\section{Axial Anomaly}
\label{appendix:fujikawa}
In this appendix, we briefly review the axial anomaly of the continuum Schwinger model, as well as its implication that observables should be invariant under changing $\theta\to\theta +2\pi$ for $m\neq 0$ and should become $\theta$-independent for $m=0$.
For this, we first note that the continuum Lagrangian of the Schwinger model,
\begin{equation}
\label{lagrangian_density}
    \mathcal{L} = \bar{\psi}(i\slashed{\partial}-g\slashed{A}-m)\psi - \frac{1}{4}F_{\mu\nu}F^{\mu\nu} + \frac{g\theta}{4\pi}\epsilon^{\mu\nu}F_{\mu\nu},
\end{equation}
is invariant under the axial transformation $\psi \to \psi' = e^{i\gamma_5\alpha}\psi$ for vanishing fermion mass, $m = 0$. However, the quantum theory described by the partition function
\begin{align}
\label{partition_function}
    &Z[m,\theta] = \int \mathcal{D}A\mathcal{D}\bar{\psi}\mathcal{D}\psi \text{ } e^{iS[A,\bar{\psi},\psi,m,\theta]}
\end{align}
with the action $S = \int d^2x\, \mathcal{L}$
does not have this symmetry, even for $m = 0$. This is due to the axial quantum anomaly, which, following the Fujikawa method~\cite{fujikawa_method}, can be shown to result from the Jacobian $J$ of the axial transformation $\psi \to \psi' $ that changes the quantum measure $\mathcal{D}\psi$ in the path integral,
\begin{equation}
    J = \exp\left(-i\int d^2x \frac{g\alpha}{4\pi}\epsilon^{\mu\nu}F_{\mu\nu}\right).
    \label{eq:Jacobian}
\end{equation}
The transformations of $\bar{\psi}$ and $\psi$ in Eq.~\eqref{partition_function} yield the same change of the measures $\mathcal{D}\bar{\psi}$ and $\mathcal{D}\psi$, so for the full measure $\mathcal{D}A\mathcal{D}\bar{\psi}\mathcal{D}\psi$ we get the square of $J$. 
We can now show that $Z[m = 0,\theta]$ is identical to $Z[m = 0,\theta = 0]$,
\begin{align}
\begin{split}
    Z[m=0,\theta] &= \int \mathcal{D}A\mathcal{D}\bar{\psi}\mathcal{D}\psi \text{ } e^{iS[A,\bar{\psi},\psi,\theta]}\\
    &= \int \mathcal{D}A\mathcal{D}\bar{\psi}'\mathcal{D}\psi' \text{ } e^{iS[A,\bar{\psi},\psi,\theta]}\\
    &= \int J^2\mathcal{D}A\mathcal{D}\bar{\psi}\mathcal{D}\psi \text{ } e^{iS[A,\bar{\psi},\psi,\theta]}\\
    &= \int \mathcal{D}A\mathcal{D}\bar{\psi}\mathcal{D}\psi \text{ } e^{iS[A,\bar{\psi},\psi,\theta]-i\int d^2x \frac{g\alpha}{2\pi}\epsilon^{\mu\nu}F_{\mu\nu}}\label{a7}\\
    &= Z[m=0,\theta = 0],
\end{split}
\end{align}
 which implies that $\theta$ is unphysical for $m = 0$.
For the second equality in Eq.~\eqref{a7}, we used the fact that the action is invariant under the axial transformation $\psi\to\psi'$ for $m = 0$. For the last equality, we set $\alpha = \theta/2$, which cancels the $\theta$-term present in the original action. 

In the presence of a nonzero mass term in the action, we would have found that $Z[m, \theta = \pi]$ is identical to $Z[-m, \theta = 0]$ because the mass term acquires a factor of $e^{2i\gamma_5\alpha}$ under the axial transformation.

Finally, we note that the last term in the fourth line of Eq.~\eqref{a7}, excluding the prefactor $i2\alpha=i\theta$, is the topological charge, which takes only integer values~\cite{coleman_more_about_SM}. Hence, the partition function is unaffected by shifting $\theta \to \theta + 2\pi$, just as any observable of the theory. This periodicity in the $\theta$-parameter was also observed for the mass shift, as shown in Fig.~\ref{fig:mass_shift_vs_l_0}.

\section{Staggered Fermions}
\label{appendix:staggered}

In this appendix, we demonstrate that our method for computing the mass shift is not limited to the case of Wilson fermions. Indeed, we apply the method to staggered fermions, which are currently being mostly used in simulations of lattice gauge theories with tensor networks and quantum computing. 

\begin{figure}[t!]
    \centering
\includegraphics[width=\linewidth]{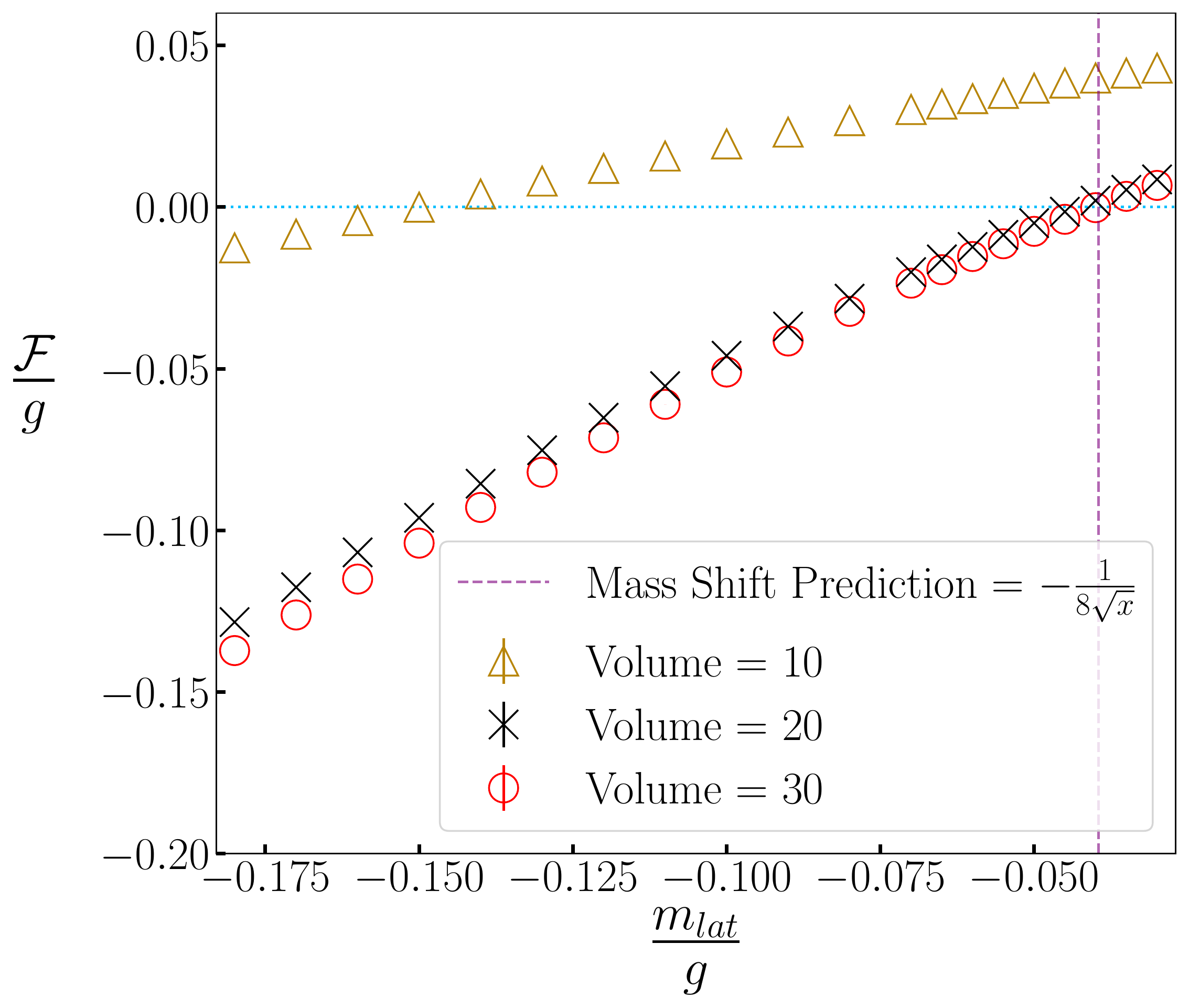}
    \caption{Electric field density $\mathcal{F}/g$ versus lattice mass $m_{\text{lat}}/g$ for staggered fermions with $x = 10$ and $l_0 = 0.125$. The blue dotted horizontal line indicates where the electric field density vanishes, and the vertical dashed purple line shows the theoretical prediction for the mass shift from Ref.~\cite{chiral_dempsey_staggered}. The markers represent data for different physical volumes $N/\sqrt{x}=10$ (yellow triangles), 20 (black crosses), and 30 (red circles). The error bars result from the extrapolation in bond dimension. As before, the error bars are much smaller than the markers and thus are not visible.}
    \label{fig:staggered}
\end{figure}

Recently, Ref.~\cite{chiral_dempsey_staggered} derived an analytical prediction for the additive mass renormalization of staggered fermions using a system with PBC. This derivation was based on enforcing a discrete spurious chiral symmetry given by a translation of one lattice site followed by shifting $\theta$ by $\pi$. The resulting mass shift is given by
\begin{align}
    \frac{m_r}{g} = \frac{m_\text{lat}}{g} + \frac{1}{8\sqrt{x}}.
    \label{eq:renormalized_mass_staggered}
\end{align}

Using the same approach as we discussed in the main text, we can numerically compute this mass shift by identifying the point at which the electric field density vanishes with $m_r/g=0$. Figure~\ref{fig:staggered} shows our numerical data for the electric field density using staggered fermions and OBC. For small volumes, we observe a noticeable difference from Eq.~\eqref{eq:renormalized_mass_staggered}, which is expected due to the different boundary conditions in our simulations. As we increase the volume, the boundary conditions become less important and the data eventually converge to the theoretical prediction.

\bibliography{References}

\end{document}